\documentclass[oldversion]{aa}  
\usepackage{graphicx}
\usepackage{txfonts}
\usepackage[authoryear]{natbib}
\bibpunct{(}{)}{;}{a}{}{,} % to follow the A&A style
\newcommand{\cotwo}{$^{12}$CO(2-1)}

\newcommand{\ea}{et al.}

\newcommand{\kms}{\>{\rm km}\,{\rm s}^{-1}}

\newcommand{\pc}{\>{\rm pc}}

\newcommand{\mpc}{\>{\rm Mpc}}
\newcommand{\m}{\>{\rm m}}

\newcommand{\mm}{\>{\rm mm}}

\newcommand{\msun}{\>{\rm M_{\odot}}}
\newcommand{\msunyr}{\>{\rm M_{\odot}\,yr^{-1}}}

\newcommand{\dg}{^{\circ}}

\newcommand{\as}{^{\prime\prime}}

\newcommand{\bdm}{\begin{displaymath}}
\newcommand{\edm}{\end{displaymath}}
\newcommand{\beq}{\begin{equation}}
\newcommand{\eeq}{\end{equation}}
\newcommand{\bit}{\begin{itemize}}
\newcommand{\eit}{\end{itemize}}
\newcommand{\ben}{\begin{enumerate}}
\newcommand{\een}{\end{enumerate}}
\newcommand{\bfi}{\begin{figure}[htb]}
\newcommand{\bpfi}{\begin{figure}[p]}

\newcommand{\paa}{$\rm Pa\alpha$}

\newcommand{\jybkms}{$\rm Jy\,beam^{-1}km\,s^{-1}$}

\begin{document}
\title{Bar-Driven Mass Build-Up within the Central 50\,pc of NGC\,6946
\thanks{Based on observations carried out with the IRAM Plateau de Bure 
Interferometer. IRAM is supported by INSU/CNRS (France), MPG (Germany) 
and IGN (Spain).}}

%   \subtitle{I. Overviewing the $\kappa$-mechanism}

   \author{E. Schinnerer\inst{1}
           \and T. B\"oker\inst{2}
           \and E. Emsellem\inst{3}
           \and D. Downes\inst{4}
          }

%   \offprints{G. Wuchterl}

   \institute{Max-Planck-Institut f\"ur Astronomie, K\"onigstuhl 17, 
              D-69117 Heidelberg, Germany\\
              \email{schinner@mpia.de}
          \and 
             European Space Agency, Dept. RSSD, Keplerlaan 1, 2200 AG 
             Noordwijk, Netherlands\\
             \email{tboeker@rssd.esa.int}
          \and 
             CRAL-Observatoire, 9 avenue Charles Andr\'e, 69231 Saint 
             Genis Laval, France\\
             \email{emsellem@obs.univ-lyon1.fr}
          \and
              Institut de RadioAstronomie Millim\'{e}trique, 300 rue de la 
              Piscine, Domaine Universitaire, 38406 Saint Martin d'H\`{e}res, 
              France
              \email{downes@iram.fr}            }

   \date{Received 7 November 2006 / Accepted 5 December 2006}

   \titlerunning{Central 50\,pc of NGC 6946}
   \authorrunning{Schinnerer et al.}
       
   \abstract {We have used the new extended A configuration of the
     IRAM Plateau de Bure interferometer to study the dense molecular
     gas in the nucleus of the nearby spiral galaxy NGC\,6946 at
     unprecedented spatial resolution in the HCN(1-0) and \cotwo\
     lines. The gas distribution in the central 50\,pc has been
     resolved and is consistent with a gas ring or spiral driven by
     the inner 400\,pc long stellar bar. For the first time, it is
     possible to directly compare the location of (dense) giant
     molecular clouds with that of (optically) visible HII regions in
     space-based images. We use the 3\,mm continuum and the HCN
     emission to estimate in the central 50\,pc the star formation
     rates in young clusters that are still embedded in their parent
     clouds and hence are missed in optical and near-IR surveys of
     star formation. The amount of embedded star formation is about
     1.6 times as high as that measured from HII regions alone, and
     appears roughly evenly split between ongoing dust-obscured star
     formation and very young giant molecular cloud cores that are
     just beginning to form stars. The build-up of central mass seems
     to have continued over the past $\ge$ 10 Myrs, to have occurred
     in an extended (albeit small) volume around the nucleus, and to
     be closely related to the presence of an inner bar.  }
   %The HCN line luminosity and the 3\,mm free-free continuum emission each
   %indicate a star formation rate of 0.1 $\msunyr$
   %, and the 6cm non-thermal
   %continuum all indicate a star formation rate of 0.1 $\msunyr$
   %likely sustained over the past $\le$ 10 Myrs.


   \maketitle
%
%________________________________________________________________

\section{Introduction}

The nearby (D=5.5$\mpc$, $1\as$=27$\pc$) Scd spiral galaxy NGC\,6946
is an ideal laboratory to study the causes (and consequences) of
intense star formation in the vicinity of a galaxy nucleus, although
its low galactic latitude hampers studies at optical and shorter
wavelengths. Within its central $300\pc$, NGC\,6946 is currently
undergoing an intense burst of star formation, as indicated by strong
far-infrared emission \citep{dev93} and the presence of numerous
hydrogen recombination lines in its nuclear spectrum \citep{eng96}.
% The HST/NICMOS \paa image of \cite{boe99} shows a complex and patchy
% morphology within the central which and the .  Despite its low
% inclination ($i=31.1\deg$) the exact position of the NGC\,6946
% nucleus can not be determined with any degree of certainty. It
% appears to be hidden behind
The proximity of NGC\,6946, combined with recent technological
advances in mm-interferometers which now routinely reach sub-arcsec
resolution, enables a detailed study of the interplay between
infalling dense molecular gas, active star formation, and the
energetic feedback from young massive stars on scales of individual
giant molecular clouds (GMCs).  We have recently undertaken such a
study which has shown that NGC\,6946 appears to be a ``textbook case''
for molecular gas responding to the gravitational potential of a
small-scale stellar bar \citep[][hereafter paper I]{sch06}.  The gas
flows inward along an S-shaped spiral structure and accumulates in a
massive nuclear clump with a size of about $2\as$ ($60\pc$). This
clump contains about $1.6\times 10^7\msun$ of molecular gas, and
appears to completely obscure the very center of NGC\,6946.

In this letter, we describe results from recent mm-observations
obtained with the new, expanded baselines of the IRAM Plateau de Bure
interferometer (PdBI) which yield a spatial resolution of about
$0.35\as$, and thus allow a direct comparison with the best available
(space-based) optical and/or near-infrared maps. The new PdBI observations
were designed to address a number of open issues related to the
connection between active star formation and the gas properties.
% location of the densest gas clumps. Despite the well-understood
% overall behavior of the the molecular gas flow, many details related
% to the exact correlation between active star formation and the
% location of the densest gas clumps remain open.
Of particular interest is the question how the sites of active star
formation compare to the molecular gas flow, and whether optical or
even near-infrared recombination lines yield a complete picture of the
current star formation rate in the nucleus of NGC\,6946. We briefly 
describe the observations and data reduction procedures in
\S\,\ref{sec:obs}, and present the resulting maps of the \cotwo\ and
HCN emission in \S\,\ref{sec:results}. In \S\,\ref{sec:central}, we
analyze the gas dynamics and various star formation tracers, and 
briefly discuss the implications of our results.

%__________________________________________________________________

\section{Observations and Data Reduction}\label{sec:obs}

We used the dual-receiver capability of the PdBI to obtain
simultaneous observations of the HCN(1-0) and \cotwo\ lines at $3\mm$
and $1\mm$, respectively. The 9\,hr long observations were performed
on January 15th, 2006, with all six PdBI antennas in their new,
extended A configuration providing baselines between $136\m$ and
$760\m$.  For calibration and mapping, we used the standard IRAM
GILDAS software packages CLIC and MAPPING \citep{gui00}. The phase
center of the observations was at 20h34m52.36s +60d09m15.96s
(J2000.0). All velocities are reported relative to $\rm
v_{sys}(LSR)\,=\,50\,\kms$.  During the observations, the typical
system temperature in the sideband containing the line of interest at
89 (230)\,GHz was about 110 (300)\,K.

The new \cotwo\ observations were combined %in the $uv$ plane 
with our previous dataset (presented in paper I) that was obtained with 
the same spectral setup in the old AB configuration. The data reduction 
and imaging closely followed the procedures described in paper I.
% were done similarly to the previous PdBI \coone\ and \cotwo\ data (see paper I for details). 
In order to obtain emission line-only maps, we subtracted (in the $uv$ plane)
continuum maps obtained from line-free channels over the 
following spectral ranges: 88.335--88.579\,GHz, 88.654--88.704\,GHz, and 
88.767--88.886\,GHz for the $3\mm$ band, and 230.211--230.367\,GHz and 
230.605--230.761\,GHz for the $1\mm$ band. 
%In order to obtain clean emission line maps, the continuum
% emission was subtracted in the $uv$ plane. 
Using uniform weighting, the resulting CLEAN beams are
$0.97\as\,\times\,0.65\as$ (PA $93\dg$) and $0.40\as\,\times\,0.29\as$
(PA $90\dg$) for the HCN(1-0) and \cotwo\ line, respectively. For both
lines, we defined a CLEAN region based on the 0th moment map. The rms
in the $6\kms$ wide channels is 2.3\,mJy\,beam$^{-1}$ for HCN(1-0) and
4.8\,mJy\,beam$^{-1}$ for CO(2-1).  The rms in the continuum maps is
0.2 (0.21) mJy\,beam$^{-1}$ and 0.7 mJy\,beam$^{-1}$ for natural
(uniform) weighting at $3\mm$ and $1\mm$, respectively. Moment maps
were calculated using the GIPSY task 'MOMENTS' requiring that emission
is present above a 3(2)$\sigma$ clipping level in at least two
adjacent channels of the CO (HCN) data cube.

%__________________________________________________________________

\section{Results}\label{sec:results}

\begin{figure}
\includegraphics[angle=0,scale=.48]{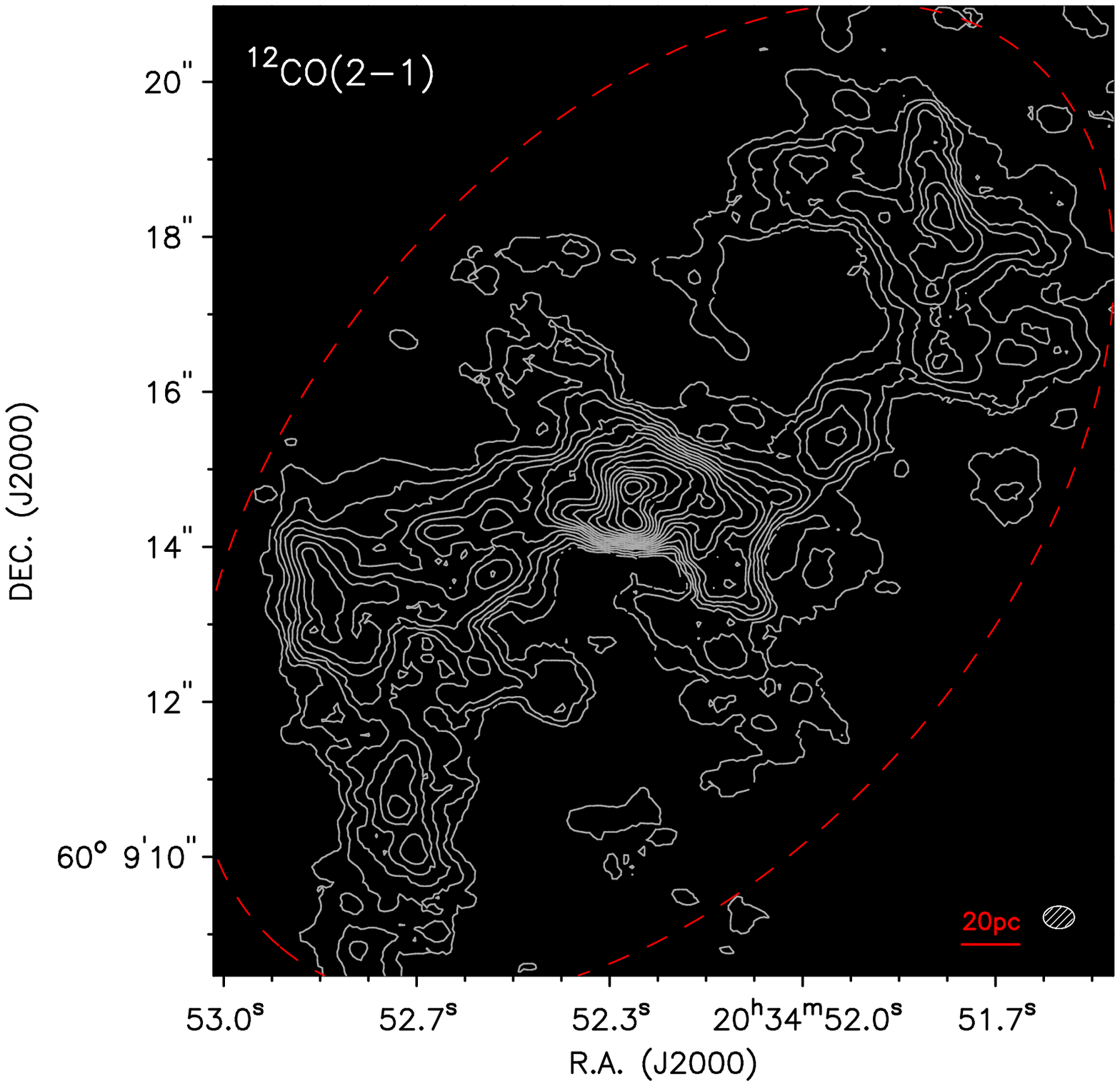}
\caption{Intensity map of the \cotwo\ line emission at
  0.4''$\times$0.3'' resolution (color and contours). The black cross
  marks the position of the dynamical center as derived in paper I.
  The beam size and the linear scale are shown for reference in the
  bottom right corner. The contours start at 0.6 \jybkms\ with the
  same step size. The red dashed ellipse indicates the location
    of the assumed bar (see paper I).
\label{fig:co}
}
\end{figure}

\begin{figure}
\includegraphics[angle=0,scale=.48]{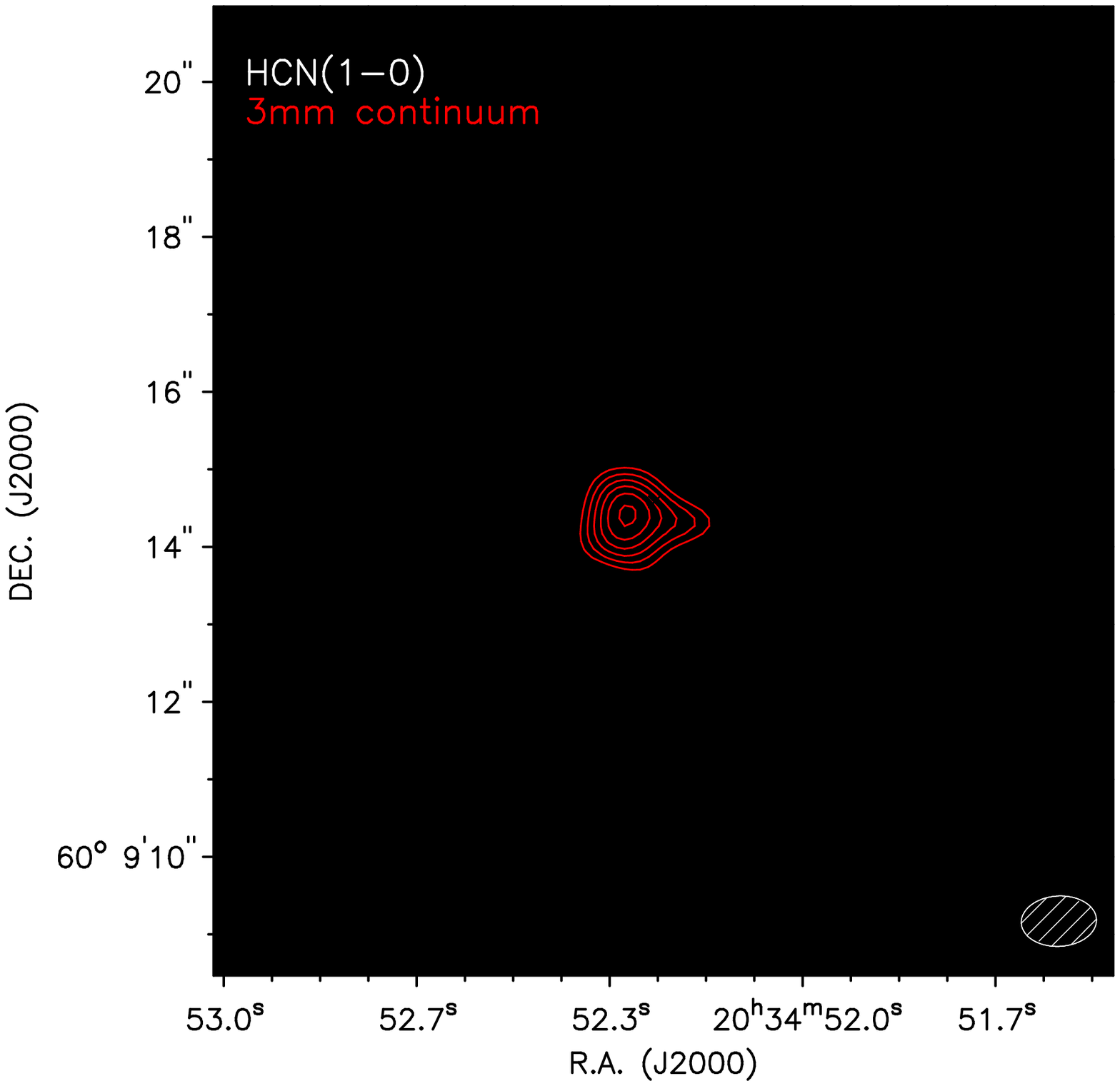}
\caption{Comparison of the intensity map of the HCN(1-0) line emission
(color) and the neighboring continuum emission (contours) at 
1.0''$\times$0.7'' resolution. The contours start at 3$\sigma$ and are in 
steps of 1$\sigma$=0.21 mJy\,beam$^{-1}$. The black cross marks the position of the 
dynamical center as derived in paper I. The beam size is shown in the bottom 
right corner.
\label{fig:hcn}
}
\end{figure}

%\subsection{\cotwo\ Emission}
\subsection{Distribution and Kinematics of CO}

As evident from Fig.\,\ref{fig:co}, the new %0.35$\as$ resolution 
\cotwo\ observations clearly resolve the
structure of the inner CO spiral\footnote{For the description of the 
various morphological components, we adopt the nomenclature 
introduced in paper I.} described in paper I. 
The south-eastern %straight 
gas lane splits into two components while the
north-western lane still appears as a single structure. More
importantly, the structure of the nuclear clump is resolved into two
%bright 
peaks north and south of our derived dynamical center with
extensions to the east and west which connect to the %straight 
gas lanes. Overall, the molecular gas is in clumps that are mostly
unresolved at our resolution of (11x8)\,pc$^2$, suggesting that the
GMCs are rather compact, dense, and gravitationally bound.

\begin{figure}
\begin{center}
\includegraphics[angle=0, scale=0.24]{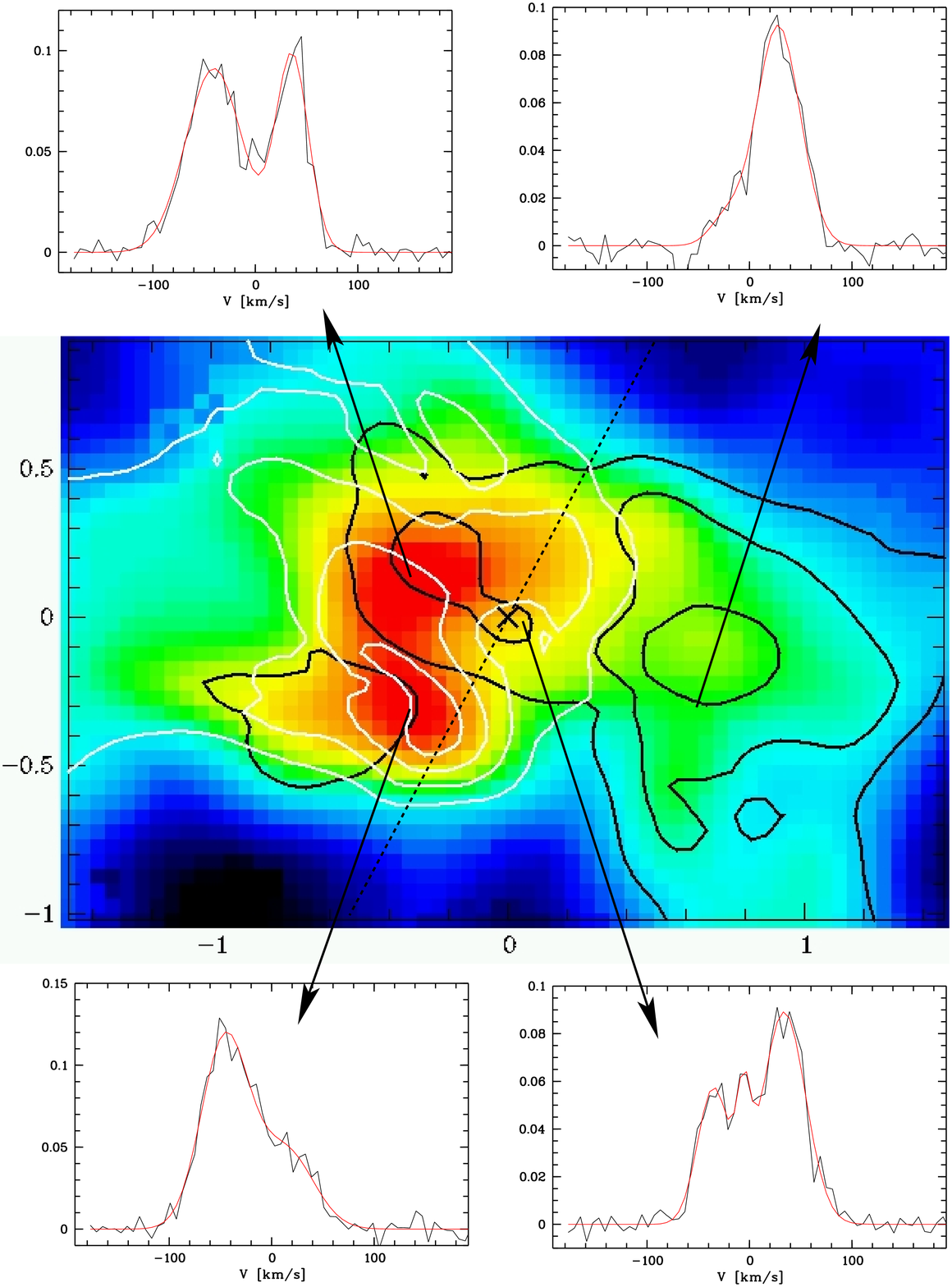}
\end{center}
\caption{Gas kinematics in the central 3$\as$ of NGC\,6946.  {\it
    Middle panel:} Integrated flux of the negative (positive) velocity
  component from a double Gaussian fit [see text; white (black)
  contours] overlaid on the total integrated \cotwo\ flux (color).
  {\it Top/Bottom panels:} Typical \cotwo\ spectra (from single
  pixels) and their corresponding Gaussian fits display a
  double-peaked profile throughout the nuclear region; the profile
  close to the dynamic center shows a third component close to the
  systemic velocity. The velocity is given relative to the systemic
  velocity and the y-axis is line intensity in Jy\,beam$^{-1}$ (with a
  temperature conversion of 200\,$\rm K\,Jy^{-1}$).}
\label{fig:fitGauss}
\end{figure}

The emission lines in the central $\sim 3\as$ have complex profiles we
fit with 2- or 3-component Gaussians. They often have two peaks, at
$(-45\pm10)\kms$ and $(+35\pm10)\kms$ relative to the systemic
velocity, with a third peak at $(-5\pm5)\kms$ in the inner 0.4'', near
the dynamical center (Fig. \ref{fig:fitGauss}). The velocity
dispersions are 20 to 30 km/s in the two main peaks, and $<\,10\,\kms$
in the third, central component. Most of the negative velocity
component is confined to one side of the dynamical minor axis, a
result which is consistent with the gas kinematics being dominated by
circular motions. This is also true for the positive velocity
component besides an extra peak about $0.3\as$ NE of the dynamical
center, close to the northern CO peak in the integrated map. This is
also illustrated by the double-peaked line profile of the northern
peak (Fig. \ref{fig:fitGauss}) when compared with the simpler line
profile of the southern peak.

\subsection{HCN(1-0) Emission}

The HCN(1-0) emission associated with the nuclear and the southern clump
is shown in Fig.\,\ref{fig:hcn}. The map contains a total flux of about
4.5\,Jy\,$\kms$ above a 2$\sigma$ threshold, that is about 8\% of the
flux of 11.4\,K\,$\kms$ measured for the central position by
\cite*{gao04} using the IRAM 30m telescope (assuming a
temperature-to-flux conversion of 4.8\,Jy\,$\rm K^{-1}$). 60\% of the
recovered flux originates in the nuclear clump. The average line
intensity ratio of HCN(1-0)/\cotwo\ is 0.02-0.03. Thus, the
non-detection of the northwestern part of the spiral structure is
probably due to our S/N limit combined with the lack of short spacings
in the HCN data. While there is a good correspondence between the HCN
and \cotwo\ emission in the southeastern spiral structure, the
integrated HCN emission has an additional peak in the
nuclear disk about $\sim$ 1$\as$ west of the dynamical center.
%from the peak of the
%CO(2-1) emission which is coinciding with the positive component (see
%Fig. \ref{fig:fitGauss} and \ref{fig:6cm}).

\subsection{Millimeter Continuum}

\begin{figure}
\begin{center}
\includegraphics[angle=0,scale=.45]{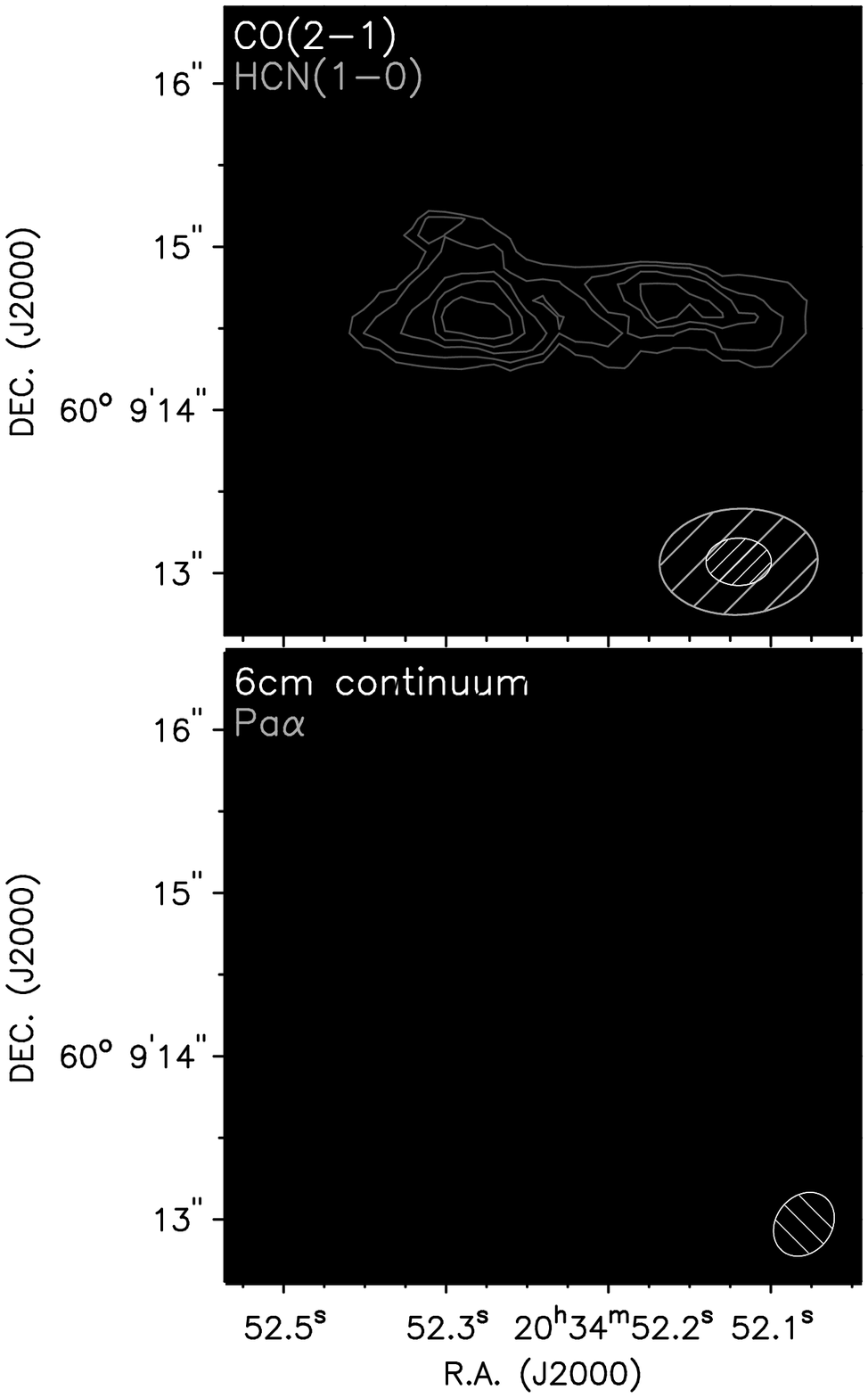}
\end{center}
\caption{{\it Top:} HCN(1-0) intensity map (contours) 
overlaid on the \cotwo\ one (color). The HCN contours start at 50\% of
the maximum in steps of 10\%. Both beams are shown in the bottom right
corner with the HCN one being the larger one.
{\it Bottom:} HST \paa\ line emission (contours) overlaid on the VLA
6cm radio continuum (color). 
%Note the remarkable similarity in the
%large scale extent as well as the details. 
The position of the dynamical center (cross) and the 3mm continuum peak 
(star) are indicated. The 6cm beam is shown in the bottom right corner.
\label{fig:6cm}
}
\end{figure}

Continuum emission has been detected at $3\mm$ and, tentatively, also
at $1\mm$.  The integrated $3\mm$ flux is $\sim$ 2.8\,mJy, with a peak
flux of 1.9 (1.7)\,mJy\,beam$^{-1}$ corresponding to a 9(8)$\sigma$
detection with natural (uniform) weighting. The position of the peak
is at 20:34:52.307 +60:09:14.37 (J2000) and offset by $\sim$ 0.4$\as$
from the dynamical center as derived in paper I. At the low levels the
$3\mm$ continuum shows an extension to the west (Fig.  \ref{fig:hcn})
and is slightly resolved in the uniformly weighted map with a
Gauss-fit of size (FWHM) $1.22\as \times 1.06\as$ (PA $\sim
96^{\circ}$). Continuum emission at $1\mm$ is present at the 4$\sigma$
level with a peak flux of 3.0\,mJy\,beam$^{-1}$ in the naturally
weighted data. A recent 0.3$\as$ resolution 6\,cm continuum map by
\cite*{tsa06} shows that the nuclear radio continuum breaks up into
two bright peaks with extensions to the north and west. The $3\mm$
continuum peak is off-set by $\sim 0.4\as$ to the north of the
bright 6\,cm peak (Fig. \ref{fig:6cm}) strongly indicating that the
3\,mm and 6\,cm continuum emission have a different origin.

%__________________________________________________________________

\section{The Central 100\,pc: Gas Flow and Star Formation}\label{sec:central}

\subsection{Bar-Driven Gas Kinematics}

In paper I, we showed that the CO spiral can be explained by the
response of the gas flow to the gravitational potential of the inner
NIR bar in the central 300\,pc. Our qualitative model requires the
existence of an Inner Lindblad resonance (ILR) at a radius of $\sim
2.2\as$ due to a $\sim$400\,pc long secondary bar. In this context, we
expect gas to accumulate at or close to the ILR and, thus, star
formation to occur. Such a mechanism is evoked to explain the
formation of the nuclear star formation rings of kiloparsec diameter
in barred galaxies \citep*[e.g. ][]{wad92,all06}. While the CO data of
paper I showed streaming motions associated with the bar, the exact
distribution of the gas in the nuclear clump could not be resolved.

The CO(2-1) intensity distribution is very reminiscent of the
so-called 'twin peaks' structure seen in lower resolution CO(1-0)
images of barred galaxies \citep*{ken92} and associated with gas
accumulation at the ILR. The two peaks of the negative and positive
velocity component are located at a radius of $\sim 0.5\as$ (Fig.
\ref{fig:fitGauss}), i.e. within the ILR of our model. Dense gas as
traced by the HCN(1-0) line roughly coincides with those CO(2-1) peaks
(Fig. \ref{fig:6cm}). One way to interpret these gas peaks is as the
contact points of two spiral arms connecting in the central 100pc. Our
derived line intensity ratios (using the integrated flux and
correcting for different beam sizes) correspond to temperature ratios
of $\rm \frac{T(HCN)}{T(CO(1-0))} \sim 0.15 - 0.20$, similar to values
found in kpc-sized starburst rings \citep*[e.g.][]{koh99}. As the
presence and distribution of dense gas is analogous to what is
observed in kpc-sized starburst rings of large-scale bars, our new
data together with the kinematic analysis (presented in detail in
paper I) are consistent with the interpretation that the inner bar
drives the CO spiral even within the nuclear clump,

The nature of the additional eastern peak in the positive velocity
component of the CO(2-1) emission is not clear. However, the young
massive stars present there (Fig. \ref{fig:6cm}) will clearly alter
the properties of the (dense) molecular gas. In addition, we start to
probe size scales where the size and stability of the GMCs themselves
is becoming important. Observations of other molecular gas tracers
might therefore be needed to clarify the situation.

\subsection{Census of Nuclear Star Formation}\label{subsec:sf}

The high extinction present in the central 100\,pc makes it difficult
to estimate the true star formation rate (SFR). In what follows, we
compare different SFR indicators in the central 3$\as \times$3$\as$ to
develop a sense for how much star formation is still deeply embedded
in dense gas clouds and hence does not reveal itself in observable
emission from hydrogen recombination lines.

The 'visible' SFR can be traced by the NIR Pa$\alpha$ line emission
arising from HII regions. Using a Pa$\alpha$ flux of $\rm
2.9\,\times\,10^{-13}\,erg\,s^{-1}\,cm^{-2}$ from the image by
\cite*{boe99} and assuming an extinction of A$_V$ = 4.6mag
\citep*{qui01}, we derive a SFR of 0.13 $\msunyr$ (see paper I for
details). This value can be directly compared to the SFR rate derived
from the non-thermal radio continuum which is associated with
synchrotron emission from supernovae explosions
\citep*[e.g.][]{con92}. \cite*{tur83} find that most ($\sim 90\%$)
emission at 6cm in the nuclear region of NGC\,6946 is
non-thermal in origin, while the detailed work by \cite*{tsa06} show
that point-like sources (either due to HII regions or supernova
remnants) contribute about 30\% to the total 6cm flux. We measure a
flux density of about 15 mJy in the central 100\,pc. Using equation
[21] of \cite*{con92} and assuming that 80\% is non-thermal in origin,
we derive a SFR of 0.1 $\msunyr$ (also taking into account stars below
5$\msun$) in agreement with the SFR derived from the Pa$\alpha$ line.
This agreement might be expected, as the spatial distributions of
Pa$\alpha$ and 6\,cm radio continuum agree very well (Fig.
\ref{fig:6cm}), and it supports our assumption that both are tracers
of stellar populations of similar age.

The millimeter continuum emission, on the other hand, traces free-free
emission from star formation sites that are still embedded in their
parent clouds. This picture is supported by two arguments: first, the
distribution of the mm continuum emission differs from that of the
radio continuum. Secondly, the spectral index $\alpha_{3-1}$ between
the 3\,mm and 1\,mm continuum flux is about 0.1 (within the
uncertainties on the 1mm flux), which is expected for free-free
emission. Following the equations summarized in \cite*{joh04} we find
an ionizing luminosity $Q_{LYC}$ of $\rm \sim 10^{52}\,s^{-1}$,
equivalent to a SFR of 0.1 $\msunyr$. This is an additional
contribution to the total SFR, originating in regions which are not
(yet) seen in hydrogen recombination lines.

Finally, \cite*{gao04} showed that there is a linear relation between
the luminosity of the HCN(1-0) line and the far-infrared luminosity.
This relation appears to not only hold true for entire galaxies, but
also for individual star forming cores within our Galaxy
\citep*{wu05}. Thus, it seems reasonable to use the HCN(1-0) line flux
as a proxy to estimate the rate of stars that are in the process of
forming. Our measured HCN(1-0) line flux of 2.5\,Jy\,km\,s$^{-1}$
corresponds to a luminosity of L'(HCN)=$\rm
3.1\times10^5\,K\,km\,s^{-1}pc^2$. Following eq. [11] of
\cite*{gao04}, we obtain another contribution of 0.06 $\msunyr$ to the
total SFR.

In summary, the different diagnostics (Pa$\alpha$/6cm, 3mm, HCN) all
trace star formation, but in different evolutionary stages. This is
also apparent from their different spatial distributions. We find
about twice as much embedded star formation (traced by HCN and 3\,mm
continuum) as star formation which has already emerged from its dust
cocoon (traced by Pa$\alpha$). This implies that
there has been significant massive star formation in the central
100\,pc of NGC\,6946 over the past $>$10 Myr. The 'older' star
formation has neither disrupted the GMCs seen in HCN nor has it
prevented the nuclear star formation evident in the 3\,mm continuum.

\subsection{Nuclear Cluster Formation?}\label{sec:disc}
             
The nucleus of most late-type spiral galaxies is marked by a compact
nuclear star cluster \citep*{boe02}. In NGC\,6946, however, such a
nuclear cluster cannot be identified. Possibly, a nuclear cluster
exists, but is obscured by the large amounts of molecular gas and dust
in the central 1". Another possibility is that we currently witness
the birth of a nuclear cluster. According to our model (paper I), the
bar-driven CO spiral structure requires the presence of an ILR, which
itself implies a mass concentration on scales of about 20-30\,pc
($\sim 1\as$). Comparison of the different star formation tracers
shows that stars are being formed in the central 60\,pc at a rate of
about 0.1\,$\msunyr$ over the past $>$ 10\,Myrs translating into a
total of roughly 10$^6\,\msun$ in stellar mass. The spatial
distribution of the SFR tracers shows that star formation occurs in
distinct areas (clusters, GMCs) spread over the central $2\as$
($60\pc$). This suggests that the build-up of central mass takes place
in an extended (albeit small) volume around the nucleus, rather than
at the very center. This is consistent with the recent identification
of blue, extended disks around a number of nuclear clusters
\citep*{set06}.
%Thus the younger stars formed must be accreted by the central mass itself. 
NGC\,6946 offers the unique opportunity to study the
processes linked to the build-up of a central mass concentration.

\begin{acknowledgements}
  ES would like to thank Philippe Salome for his help with the IRAM
  data reduction. We also thank Chao-Wei Tsai and Jean Turner for
  providing their 6cm VLA data for comparison and the referee Jonathan
  Braine for helpful comments.
\end{acknowledgements}

\end{document}